\documentclass[showpacs,aps,prd]{revtex4}
\input epsf

\begin{document}

\title{Possibility of $\Sigma_{c}(2800)$ and $\Lambda_{c}(2940)^{+}$ as $S$-wave $D^{(*)}N$ molecular states}
\author{Jian-Rong Zhang}
\affiliation{Department of Physics, College of Science, National University of Defense Technology,
Changsha 410073, Hunan, People's Republic of China}

\begin{abstract}
In this talk, we
give a brief review of our work
studying the possibility of $\Sigma_{c}(2800)$ and $\Lambda_{c}(2940)^{+}$ as $S$-wave $D^{(*)}N$ molecular states
from QCD sum rules.
\end{abstract}
\maketitle

In 2005, Belle Collaboration observed
an isotriplet of new states $\Sigma_{c}(2800)$ \cite{2800-belle},
the neutral one of which was confirmed by Babar Collaboration  \cite{2800-babar}.
Moreover, Babar collaboration reported
the observation of $\Lambda_{c}(2940)^{+}$ \cite{2940-babar},
which was confirmed by Belle Collaboration \cite{2940-belle}.

There have appeared two different ways to understand the inner structures of $\Sigma_{c}(2800)$ and $\Lambda_{c}(2940)^{+}$.
From a potential model prediction,
their masses
are close to theoretical values of $\Sigma_{c}^{*}$ with $J^{P}=\frac{3}{2}^{-}$ or $\frac{5}{2}^{-}$ and $\Lambda_{c}^{*}$ with $J^{P}=\frac{5}{2}^{-}$ or $\frac{3}{2}^{+}$, respectively \cite{quark-model}.
One way of studies grounds on the
assignments of them as conventional charmed baryons \cite{Ebert,Cheng,Chen,Zhong,Garcilazo,BChen,JHe}.
Another way bases on the assumption that they are some molecular candidates \cite{Lutz,Tejero,Dong,He,Dong1,JHe2,Recio}.
In particular, Ortega {\it et al.} studied $\Lambda_{c}(2940)^{+}$ as a $D^{*}N$ molecule
with $J^{P}=\frac{3}{2}^{-}$ in a constituent quark model, and claimed
obtaining a mass which agrees with the experimental data \cite{Ortega}.

To investigate the possibility of $\Sigma_{c}(2800)$ and $\Lambda_{c}(2940)^{+}$
as the $S$-wave $DN$ state with $J^{P}=\frac{1}{2}^{-}$ and the $S$-wave $D^{*}N$ state with $J^{P}=\frac{3}{2}^{-}$
respectively,
we devoted to studying them in Ref. \cite{ZJR} employing the method of
QCD sum rules \cite{svzsum} (for reviews
see \cite{overview1,overview2,overview3,overview4,XYZ} and references
therein; Particularly,
many theorists began to study
light pentaquark states in \cite{pentaquark,pentaquark1} and heavy ones in
\cite{pentaquark-heavy}).

The basic point of QCD sum rules is to construct a proper
interpolating current to represent the studied state.
At present, molecular currents are built up with the color-singlet currents of
composed hadrons to form hadron-hadron
configurations of fields.
One could find meson currents in Ref. \cite{reinders} and
nucleon ones in \cite{baryon-current}.
Therefore, we build following forms of currents:
\begin{eqnarray}
j&=&(\bar{q}^{c^{'}}i\gamma_{5}Q^{c^{'}})(\varepsilon_{abc}q_{1}^{Ta}C\gamma_{\mu}q_{2}^{b}\gamma_{5}\gamma^{\mu}q_{3}^{c}),
\end{eqnarray}
for the $S$-wave $DN$ or $\bar{B}N$ molecular state with $J^{P}=\frac{1}{2}^{-}$, and
\begin{eqnarray}
j^{\rho}&=&(\bar{q}^{c^{'}}\gamma^{\rho}Q^{c^{'}})(\varepsilon_{abc}q_{1}^{Ta}C\gamma_{\mu}q_{2}^{b}\gamma_{5}\gamma^{\mu}q_{3}^{c}),
\end{eqnarray}
for the $S$-wave
 $D^{*}N$ or $\bar{B}^{*}N$ molecular state with $J^{P}=\frac{3}{2}^{-}$.
Here $Q$ is heavy quark $c$ or $b$, and $q_{1}$, $q_{2}$, and $q_{3}$ denote light quarks $u$ and/or $d$.
The index $T$ means matrix
transposition, $C$ is the charge conjugation matrix,
with $a$, $b$, $c$ and $c'$ as color indices.

We derive QCD
sum rules for molecular states with similar techniques
as our previous works on heavy baryons \cite{Zhang} and molecular states \cite{Zhang1}.
Ultimately,
one can gain mass sum rules \cite{ZJR}
\begin{eqnarray}\label{sumrule1}
M_{H}^{2}&=&\bigg\{\int_{m_{Q}^{2}}^{s_{0}}ds\rho_{1}(s)s
e^{-s/M^{2}}+d/d(-\frac{1}{M^{2}})\hat{B}\Pi_{1}^{\mbox{cond}}\bigg\}/
\bigg\{\int_{m_{Q}^{2}}^{s_{0}}ds\rho_{1}(s)e^{-s/M^{2}}
+\hat{B}\Pi_{1}^{\mbox{cond}}\bigg\},
\end{eqnarray}
where $M_{H}$ is the mass of the hadronic resonance, $s_0$ is the threshold parameter,
 and $M^2$ indicates the Borel parameter.
The spectral density $\rho_{1}(s)$ is given by the correlator's imaginary part.
In its calculation, one works at leading order in $\alpha_{s}$ and considers condensates up
to dimension $12$.
To keep the heavy-quark mass finite, one can use the
momentum-space expression for the heavy-quark propagator \cite{reinders}.
The light-quark part of the
correlator can be calculated in the coordinate space employing the light-quark
propagator,
which is then
Fourier-transformed to the momentum space in $D$ dimension.
The
resulting light-quark part is combined with the heavy-quark part
before it is dimensionally regularized at $D=4$.

Numerically, one could take input parameters as
$m_{c}=1.23\pm0.05~\mbox{GeV}$, $m_{b}=4.24\pm0.06~\mbox{GeV}$,
$\langle\bar{q}q\rangle=-(0.23\pm0.03)^{3}~\mbox{GeV}^{3}$, $\langle
g\bar{q}\sigma\cdot G q\rangle=m_{0}^{2}~\langle\bar{q}q\rangle$,
$m_{0}^{2}=0.8\pm0.1~\mbox{GeV}^{2}$, $\langle
g^{2}G^{2}\rangle=0.88~\mbox{GeV}^{4}$, and $\langle
g^{3}G^{3}\rangle=0.045~\mbox{GeV}^{6}$
\cite{overview2}.
For the present case, it may have some difficulty to find a standard work window critically satisfying all
the conventional rules, which has been discussed in detail for some other cases \cite{HXChen,ZGWang,Matheus,Zs}.
By releasing the rigid OPE convergence criterion, one could choose some transition range as a compromise Borel window
and arrive at
$3.75\pm0.14\pm0.08~\mbox{GeV}$
for the $S$-wave $DN$ state with $J^{P}=\frac{1}{2}^{-}$.
To investigate the effect of different factorization of $\langle q\bar{q}q\bar{q}\rangle=\kappa\langle\bar{q}q\rangle\langle\bar{q}q\rangle$, we average three results
for $\kappa=1\sim3$ and arrive at the mass value $3.64\pm0.33~\mbox{GeV}$ concisely for the
$S$-wave $DN$ state with $J^{P}=\frac{1}{2}^{-}$,
which is somewhat higher than the experimental value of $\Sigma_{c}(2800)$ even considering the result's uncertainty.
For the $S$-wave $\bar{B}N$ state with $J^{P}=\frac{1}{2}^{-}$,
one can obtain the mass value $6.97\pm0.34~\mbox{GeV}$.
Similarly, the final result is
$3.73\pm0.35~\mbox{GeV}$ for the
$S$-wave $D^{*}N$ state with $J^{P}=\frac{3}{2}^{-}$,
which is greater than the experimental data of $\Lambda_{c}(2940)^{+}$
even taking into account the uncertainty.
For the mass of $S$-wave $\bar{B}^{*}N$ state with $J^{P}=\frac{3}{2}^{-}$,
one could gain
$6.98\pm0.34~\mbox{GeV}$.

In summary, we have studied the possibility of $\Sigma_{c}(2800)$ and $\Lambda_{c}(2940)^{+}$
as $S$-wave $D^{(*)}N$ states using QCD sum rules in Ref. \cite{ZJR}. The final results are respectively greater than the experimental data of $\Sigma_{c}(2800)$ and $\Lambda_{c}(2940)^{+}$.
In view of that corresponding molecular currents are constructed from local operators
of hadrons, the possibility of $\Sigma_c(2800)$ and $\Lambda_{c}(2940)^{+}$ as molecular states
can not be arbitrarily excluded merely from these disagreements
between molecular masses using local currents and experimental data.
However, one can infer that $\Sigma_{c}(2800)$ and $\Lambda_{c}(2940)^{+}$ could not be compact states from these results.
This may suggest a limitation of the QCD sum rule
using the local current to determine whether some state is a molecular state or not.
As byproducts, masses for their
bottom partners are predicted to be $6.97\pm0.34~\mbox{GeV}$ for the $S$-wave $\bar{B}N$ state of $J^{P}=\frac{1}{2}^{-}$ and
$6.98\pm0.34~\mbox{GeV}$ for the $S$-wave $\bar{B}^{*}N$ state of $J^{P}=\frac{3}{2}^{-}$.
\begin{acknowledgments}
This work was supported by the National Natural Science
Foundation of China under Contract Nos. 11105223, 10947016, 11275268, the
Foundation of NUDT (No. JC11-02-12), and the
project in NUDT for excellent youth talents.
\end{acknowledgments}

\end{document}